\documentclass[conference]{IEEEtran}
\IEEEoverridecommandlockouts
\usepackage{cite}
\usepackage{amsmath,amssymb,amsfonts}
\usepackage{algorithmic}
\usepackage{graphicx}
\usepackage{textcomp}
\usepackage{xcolor}
\def\BibTeX{{\rm B\kern-.05em{\sc i\kern-.025em b}\kern-.08em
    T\kern-.1667em\lower.7ex\hbox{E}\kern-.125emX}}
\begin{document}
\title{A Light Weight Neural Network for Automatic
Modulation Classification in OFDM Systems\\
{\footnotesize \textsuperscript{}}
\thanks{}
}
\author{\IEEEauthorblockN{1\textsuperscript{st} Indiwara Nanayakkara}
\IEEEauthorblockA{\textit{Department of EEE} \\
\textit{University of Peradeniya  }\\
Peradeniya, Sri Lanka \\
indiwarananayakkara1@gmail.com}
\and
\IEEEauthorblockN{2\textsuperscript{nd} Dehan Jayawickrama}
\IEEEauthorblockA{\textit{Department of EEE} \\
\textit{University of Peradeniya  }\\
Peradeniya, Sri Lanka \\
dehanjayawickrama@gmail.com}
\and
\IEEEauthorblockN{3\textsuperscript{rd} Dasuni Jayawardena}
\IEEEauthorblockA{\textit{Department of EEE} \\
\textit{University of Peradeniya  }\\
Peradeniya, Sri Lanka \\
dasuj.mc@gmail.com}

\and
\hspace{4cm}
\IEEEauthorblockN{4\textsuperscript{th} Vijitha R. Herath}
\hspace{5cm}
\IEEEauthorblockA{
\hspace{4cm}\textit{Department of EEE}\\
\hspace{4cm}\textit{University of Peradeniya}\\
\hspace{4cm}Peradeniya, Sri Lanka\\
\hspace{4cm}vrherath@gmail.com
}
\and
\IEEEauthorblockN{5\textsuperscript{th} Arjuna Madanayake}
\IEEEauthorblockA{\textit{Department of ECE} \\
\textit{Florida International University}\\
Miami, Florida \\
amadanay@fiu.edu}

}

\maketitle

\begin{abstract}
Automatic Modulation Classification (AMC) is a vital component in the development of intelligent and adaptive transceivers for future wireless communication systems. Existing statistically-based blind modulation classification methods for Orthogonal Frequency Division Multiplexing (OFDM) often fail to achieve the required accuracy and performance. Consequently, the modulation classification research community has shifted its focus toward deep learning techniques, which demonstrate promising performance, but come with increased computational complexity.
In this paper, we propose a lightweight subcarrier-based modulation classification method for OFDM systems. In the proposed approach, a selected set of subcarriers in an OFDM frame is classified first, followed by the prediction of the modulation types for the remaining subcarriers based on the initial results. A Lightweight Neural Network (LWNN) is employed to identify the initially selected set of subcarriers, and its output is fed into a Recurrent Neural Network (RNN) as an embedded vector to predict the modulation schemes of the remaining subcarriers in the OFDM frame. 
\end{abstract}


\section{Introduction}
The rapid convergence of artificial intelligence (AI) and machine learning (ML) with radio frequency (RF) wireless, aerospace, and electronic warfare systems is leading to increased autonomy and self-administration of RF systems. 
Automatic modulation classification (AMC) identifies the modulation type used in a signal and serves as a link between signal detection and demodulation. Without requiring any prior information on the signal, AMC enables devices to determine the modulation type. Consequently, AMC is commonly used in both military and civilian applications, such as electronic warfare, cognitive radios, spectrum surveillance, and spectrum management. However, detailed discrimination at the bin-level for higher order digital modulation, such as in the case of orthogonal frequency division multiplexing (OFDM) systems, is an open problem and forms the motivation for this work.

\section{Review}
Modulation classification methods are classified into likelihood-based and feature-based approaches \cite{11}. The likelihood-based method is considered the most optimal according to the detection theory. However, it is impractical in real-world scenarios due to high computational complexity and requirements for unknown parameters related to the channel and the noise statistics \cite{Fu}. In contrast, the feature-based approach employs algorithms that analyze the extracted complex features of the original signal to recognize the modulation scheme  \cite{13} \cite{14}. Despite the efficiency and robustness of processing, the feature-based method tends to be less accurate for higher-order signals. 

Our proposed AMC method is designed to classify modulation schemes in scenarios where varying modulation schemes are used across subcarriers.

\begin{figure*}[htbp]
    \centering
    \includegraphics[width=0.94\textwidth]{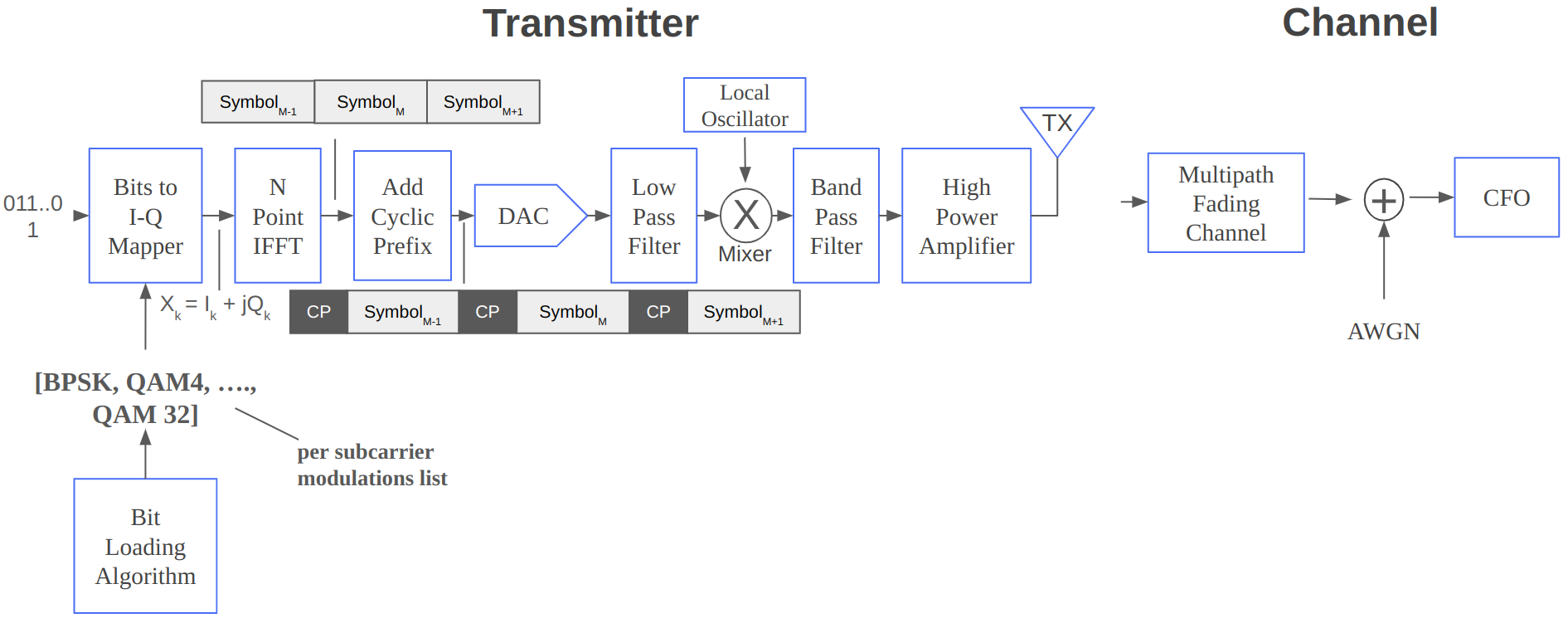}
    \caption{Typical signal flow graph of an OFDM-based digital communication system.}
    \label{fig:image_label}
    
\end{figure*}

\begin{enumerate}

    \item The proposed AMC method is optimized for environments with limited computational power, as computational efficiency was a key consideration during its design.
    \item In this work, we present a novel lightweight CNN network (LWNN), inspired by efficient architectures from the computer vision domain, designed to classify QAM4, QAM8, QAM16, QAM32, and QAM64 modulation schemes. 
    \item Additionally, we introduce a specialized approach for M-QAM OFDM systems to further reduce computational complexity with the help of RNN based classifier (RNN-BC). This method leverages the relationships between modulation schemes that result from bit and power allocation strategies.
    
\end{enumerate}

\section{AMC For OFDM SYSTEMS}

This section outlines the steps involved in OFDM modula-
tion classification, drawing on existing literature that presents
the baseband OFDM signal model and provides a general
overview of bit-loading algorithms in OFDM systems.

\subsection{OFDM Signal Model}

The overall baseband single-input single-output OFDM system is given in Fig. 1. At the transmitter,binary bits are first modulated using modulation schemes determined by the bit loading algorithm to generate OFDM symbols with N subcarriers. 

The N-point Inverse Fast Fourier Transform (IFFT) vector for the $m^{th}$ symbol in frequency domain is given in (1).

\begin{equation}
    \bar{X}_m=\left[\bar{X}_m(0), \bar{X}_m(1), \bar{X}_m(2),  \ldots, \bar{X}_m(N-1)\right]
\end{equation}

After performing the N-point FFT the time domain signal of the $m^{th}$ symbol's, $k^{th}$ sample is given in (2).

\begin{equation}
    x_m(k)=\sum_{n=0}^{N-1} \bar{X}_m(k) e^{j 2 \pi k\left(\frac{n}{N}\right)} \,\,\,\,\,\,0 < n \leq N-1
\end{equation}

After adding cyclic prefix to the $m^{th}$ symbol the $m^{th}$ symbol can be denoted by

\begin{equation}
    x_m(k)=\sum_{n=-N_{c p}}^{N-1} X_m(k) e^{j 2 \pi k\left(\frac{n}{N}\right)} \quad-N_{c p}<n \leq N-1
\end{equation}

Where \(N_{cp}\) is the cyclic prefix length.

\begin{align}
X_m(k) &= 
\begin{cases} 
\bar{X}_m(k) e^{j 2 \pi k}, & n < 0 \\
\bar{X}_m(k), & 0 \leq n \leq N - 1
\end{cases}
\end{align}

After passing through the multi-path fading channel with AWGN, final output signal of  $k^{th}$ sample of $m^{th}$ symbol is given in (4).

\begin{equation}
    s_m(k)=\sum_{n=0}^{N-1} H(k) X_m(k) e^{j 2 \pi k\left(\frac{n}{N}\right)}+ N(k) 
\end{equation}

Where
 \(N(k)\) is the AWGN at the $k^{th}$ sample, \(H(k)\) is the channel response.

\vspace{1cm}
After adding CFO and $T_{\text{o}}$ the final received signal of the $m^{th}$ symbol is given by(5).

\begin{equation}
 r_m(n)=s_m(n-\phi) e^{j 2 \pi \epsilon\left(\frac{n}{N}\right)}   
\end{equation}
where
$\epsilon$ is the normalized carrier frequency offset (CFO),  
$\phi$ is the phase offset,  
$n$ is the sample count,  
$N$ is the total number of samples,  
$r_m(n)$ is the received signal, and  
$s_m(n-\phi)$ is the transmitted signal delayed by $\phi$.

\subsection{Bit Loading and Power Allocation Methods}

In conventional wireless OFDM systems, a uniform signal constellation is employed across all subcarriers. However, the overall error probability is dictated by the subcarriers experiencing the worst channel conditions. To mitigate this issue and increase the system performance, modern OFDM systems commonly incorporate adaptive bit and power allocation algorithms. Bit and power loading algorithms can be broadly classified into two categories: margin maximization (MM) algorithms \cite{61} \cite{44} and rate maximization (RM) algorithms \cite{46} \cite{45}.

\begin{figure}[htbp]
    \centering
    \includegraphics[width=0.45\textwidth]{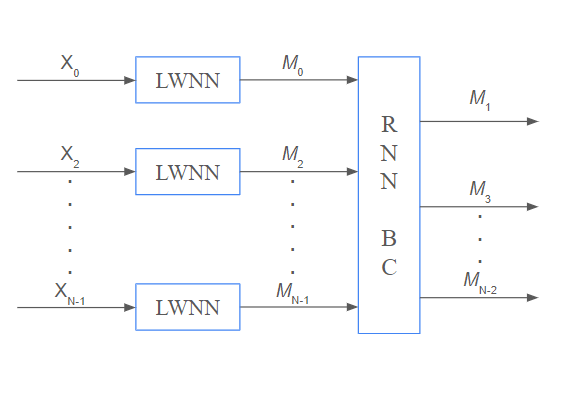}
    \caption{LWNN applied N/2 times to classify modulations in N/2 subcarriers and RNN-BC applied to the N/2 modulations classified by LWNN to predict the modulations in the remaining N/2 subcarriers.}
    \label{fig:method_2}
\end{figure}

\section{Proposed AMC Architectures}

Our proposed methodology contains two stages. In the first stage we have proposed a Light Weight Convolutional Neural Network(LWCNN/LWNN) which will identify modulation scheme in each subcarrier in the OFDM modulated signal. For an OFDM signal with N number of subcarriers the LWNN needs to be run N times to identify the modulation schemes. 

Going further with the above implementation we have proposed a way to reduce the number of times the LWNN has to run so that the computational resources are efficiently used. In the second stage we propose only to run the LWNN on N/2 number of subcarriers and get the modulation schemes of those N/2 subcarriers. Those modulation classification results are then fed into a Recurrent Neural Network(RNN) which will only run once to predict the remaining modulation schemes on N/2 subcarriers.  This is illustrated in Fig.2.

\subsection{Lightweight Neural Network Architecture}

CNNs are used in machine learning, particularly in com-
puter vision based on their ability to capture features from
feature maps regardless of spatial localization. However, CNNs are computationally expensive because of the large number of multiplications required by their design.

The proposed architecture (LWNN) incorporates elements
from \cite{49} and \cite{51} to achieve a low complexity solution
suitable for resource-constrained environments. The complete
architecture, along with layer output sizes, is shown in Fig 3.
Decomposed convolution (D-CNN) blocks (marked as D-
CNN in Fig. 3), similar to that used in \cite{49}, and inception
blocks (marked as Inception in Fig. 3), similar to that used in
\cite{51}, are used in conjunction as the main CNN components.
Architectures of D-CNN block and Inception block are shown
in Fig 5 and Fig 6 respectively.
Batch normalization layers were used after every D-CNN
block and inception block to speed up the training.The model was trained using the Adam optimizer with a learning rate of 1 $\times 10^{-3}$
, categorical cross-entropy as the loss function, and a batch size of 64.

\subsection{Recurrent Neural Network Based Classifier (RNN-BC)}

After the proposed LWNN classifies the modulations
on N/2 subcarriers, the RNN model is used to classify the
modulation schemes in the remaining subcarriers. Specifically,
the LWNN classifies the modulation schemes on all even-
numbered subcarriers, while the RNN predicts the classifi-
cations on odd-numbered subcarriers, based on the results
produced by LWNN.

In the proposed
approach, an RNN model is utilized to exploit the relationships
and correlations between modulation schemes on adjacent
subcarriers. Specifically, we utilize a bidirectional RNN model
with Gated Recurrent Units (GRUs) as RNN cells. GRUs were
chosen for their ability to effectively capture relationships in
long sequences having low computational complexity than
Long Short-Term Memory (LSTM) units.

The RNN-based classifier employs an embedding layer of dimension 32 to represent the input features, which are then processed through two stacked bidirectional GRU layers with 64 and 128 hidden units, respectively. The bidirectional configuration allows the model to capture contextual dependencies in both forward and backward directions across subcarriers. The output from the final GRU layer is passed to a fully connected classification layer for modulation prediction. The model is trained using the Adam optimizer with a learning rate of 
1 $\times 10^{-5}$ ,categorical cross-entropy as the loss function and a batch size of 32.

 \begin{figure}[htbp]
    \centering
    \includegraphics[width=0.23\textwidth]{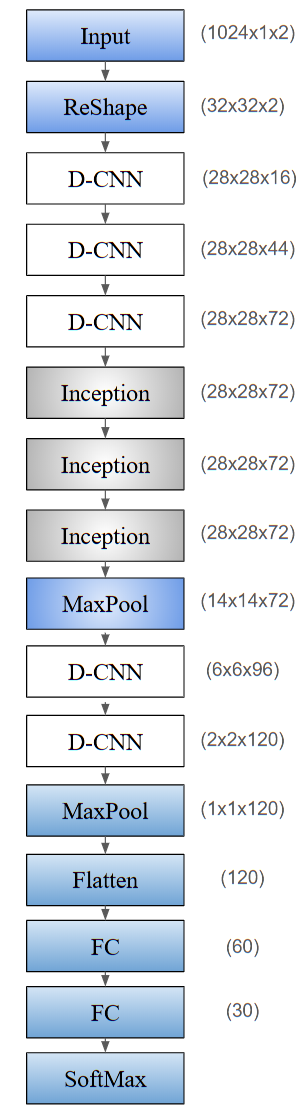}
    \caption{Backbone of the proposed lightweight neural network architecture for modulation classification}
    \label{fig:lwnn_backbone}
\end{figure}

\begin{figure}[htbp]
    \centering
    \includegraphics[width=0.48\textwidth]{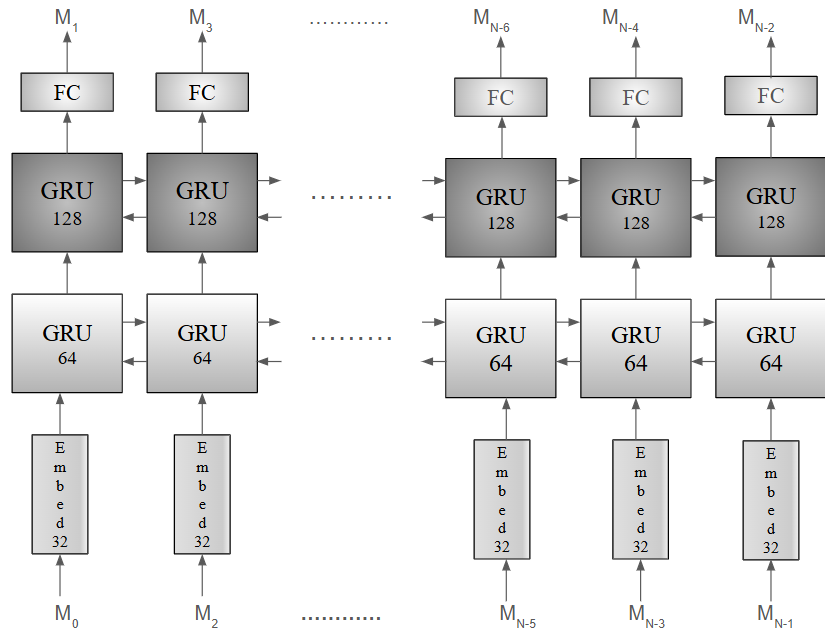}
    \caption{Architecture of the RNN-BC}
    \label{fig:rnn_bc_backbone}
\end{figure}

\begin{figure}[htbp]
    \centering
    \includegraphics[width=0.20\textwidth]{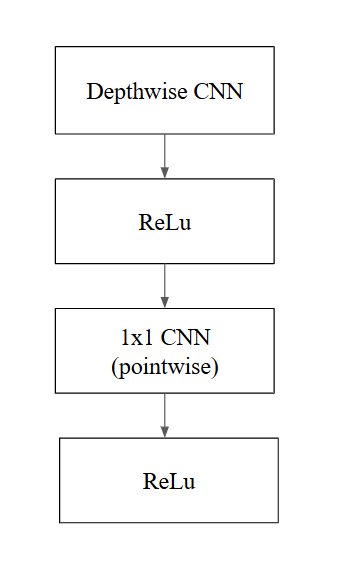}
    \caption{Architecture of Decomposed CNN (D-CNN) block employed in the proposed LWNN}
    \label{fig:d_cnn}
\end{figure}

\begin{table}[ht]
    \centering
    \begin{tabular}{|l|c|c|}
    \hline
    \textbf{Model} & \textbf{Calcu.(Model Alone)} & \textbf{Calcu.(Model+RNN-BC)} \\ \hline
    VGG                            & 16.45M                           & 8.31M                               \\ \hline
    ResNet                         & 15.1M                            & 7.63M                               \\ \hline
    CNN-AMC                        & 36.8M                            & 18.48M                              \\ \hline
    \end{tabular}
    \vspace{0.2cm} 
    \caption{Comparison of the calculations required to classify modulations in a 64-subcarrier OFDM signal.}
    \label{tab:complexity_comparison_rnnbc}
\end{table}

\begin{figure*}[htbp]
    \centering
    \includegraphics[width=0.70\textwidth]{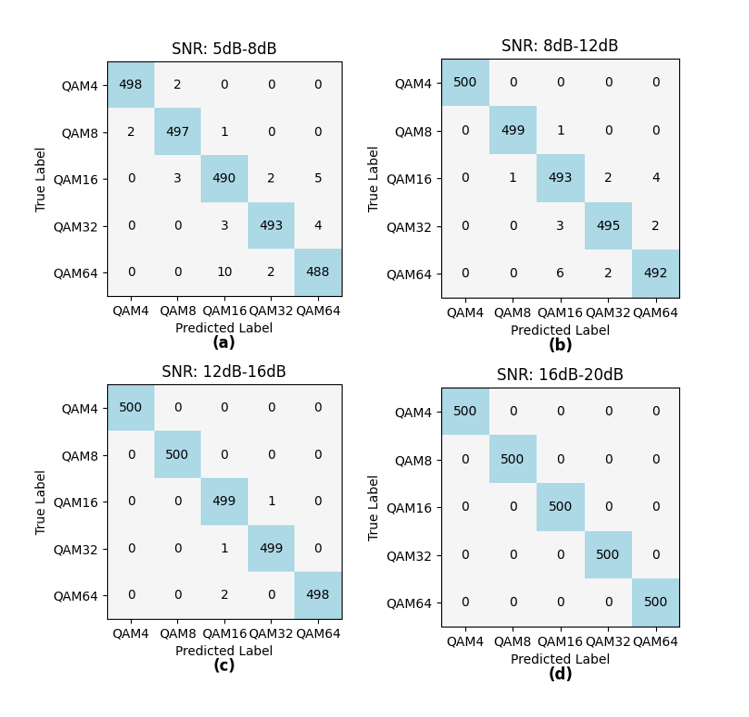}
    \caption{(a)Classification results  using the LWNN for SNR 5dB to SNR 8dB (b)Classification results  using the LWNN for SNR 8dB to SNR 12dB (c)Classification results using the LWNN for SNR 12dB to SNR 16dB
(d)Classification results using the LWNN for SNR 16dB to SNR 20dB}
    \label{fig:cf_mat}
\end{figure*}

\begin{figure}[htbp]
    \centering
    \includegraphics[width=0.39\textwidth]{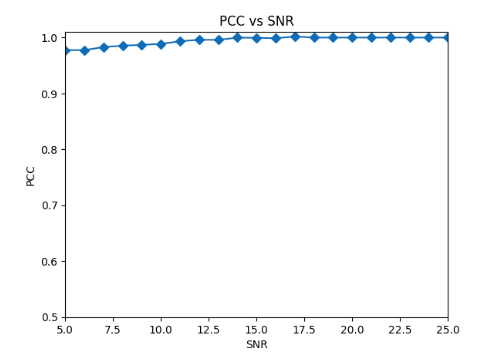}
    \caption{PCC vs SNR variation for modulation schemes QAM4, QAM8, QAM16, QAM32, QAM64 using LWNN and RNN-BC combined}
    \label{fig:pcc_vs_snr}
\end{figure}

\begin{figure}[htbp]
    \centering
    \includegraphics[width=0.32\textwidth]{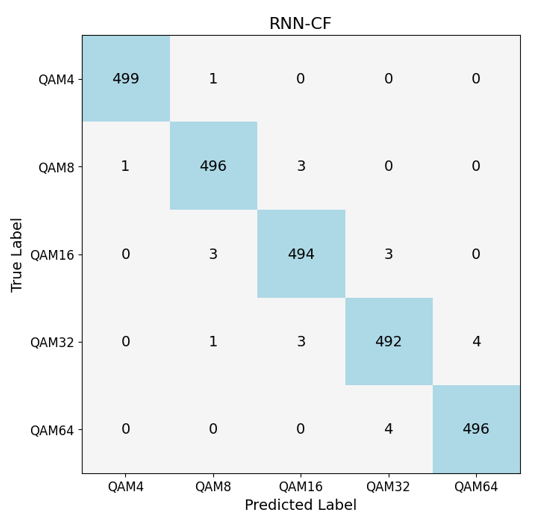}
    \caption{Classification results of RNN-BC model}
    \label{fig:rnn_bc_cf}
\end{figure}

\section{Analysis of Dataset, Results and Computational Complexity}

This section describes the datasets used for training and testing the models, and presents the results along with an analysis of the computational complexity.

\subsection{Dataset Description}

A 64-subcarrier, 20 MHz bandwidth OFDM system was considered for signal generation. Cyclic prefix length was set to 16 samples, resulting in a total symbol length of 80 samples and each signal consists of 1024 I-Q samples.
For synthetic channel conditions, the Rayleigh fading multipath channel model was used to model the channel, with the maximum delay spread set to be less than the cyclic prefix duration. The number of multipath channels was uniformly sampled between 2 and 10 during signal generation. The noise spectral density was adjusted to ensure that the SNR across subcarriers remained approximately between 5 dB and 25 dB. The minimum SNR was maintained above 5 dB, as the bit-loading algorithm nullifies subcarriers when SNR levels are too low.
    
In total, four datasets were created: two training sets for training the LWNN and RNN-BC models, and two test sets for evaluating these models. The bit-loading algorithm proposed in \cite{45} was used to generate the training and test datasets for the RNN-BC.

\subsection{CNN Based AMC Results (LWNN)}

Probability of Correct Classification (PCC) is a metric
used to evaluate the performance of a classification model,
particularly in scenarios like modulation classification. PCC
measures the proportion of correctly classified instances out of
the total number of instances.
Higher PCC values indicate better classification performance.Above 97 percent accuracy across the considered SNR ranges were obtained using the LWNN alone.  Fig.~\ref{fig:cf_mat} presents the confusion matrices plotted for different SNR ranges using the LWNN model alone. Subsequently, the same 10,000 signals were classified using
both the LWNN and RNN-BC models in conjunction. Specif-
ically, for each signal, 32 subcarriers were classified using
the LWNN model, while the remaining 32 subcarriers were
classified using the RNN-BC model. 

\begin{figure}[htbp]
    \centering
    \includegraphics[width=0.47\textwidth]{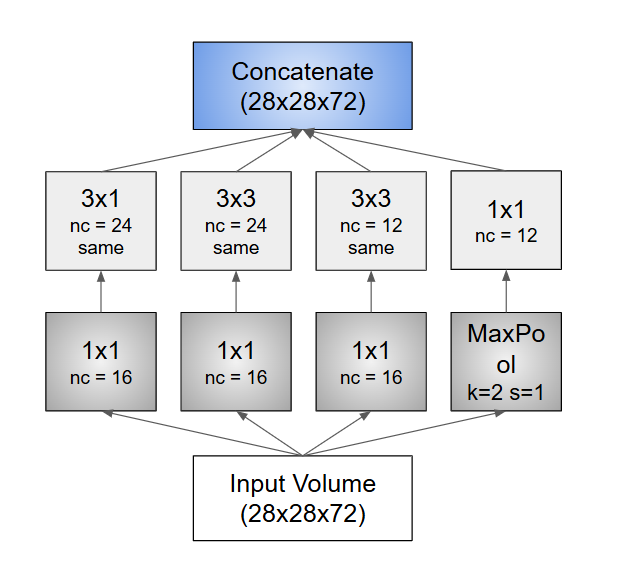}
    \caption{architecture of Inception block employed in the Proposed LWNN}
    \label{fig:inception}
\end{figure}

Subsequently, the same 10,000 signals were classified using both the LWNN and RNN-BC models in conjunction. Specifically, for each signal, 32 subcarriers were classified using the LWNN model, while the remaining 32 subcarriers were classified using the RNN-BC model. Fig.~\ref{fig:pcc_vs_snr} illustrates the variation of the PCC with respect to the SNR for the results of the 10,000 signals.

\subsection{RNN Based AMC Results (RNN-BC)}

A total of 2500 generated modulation lists, each with a
length of 64, were used to evaluate the RNN-BC model alone.
Modulations of even subcarriers are fed into the RNN-BC
model, which classifies the modulations of the remaining 32
subcarriers. Fig. 9 presents the classification results of the
RNN-BC model alone.

\subsection{Computational Complexity}

The computational complexity of a deep learning model is represented by the number of floating-point operations (FLOPs) required per inference. The proposed LWNN model requires approximately 48.9 million FLOPs, while the RNN-BC classifier requires about 75520 FLOPs. Considering a 64-subcarrier system that employs different modulation schemes across subcarriers, using only the LWNN model would require running it 64 times, resulting in roughly 64 $\times$ 48.9M $\approx$ 3.13 billion FLOPs in total.

However, when the RNN-BC is used along with the LWNN, only 32 subcarriers are needed to be classified by the LWNN, requiring 32 $\times$ 48.9M $\approx$ 1.56 billion FLOPs. Additionally, a single run of the RNN-BC model is required to classify the remaining modulations, which adds about 76000 FLOPs. This combined approach results in a total of approximately 1.56 billion FLOPs. Therefore, using the proposed LWNN and RNN-BC together nearly halves the computational complexity, compared to using a conventional CNN architecture for modulation classification alone.

 The computational complexity of a deep learning model is
directly proportional to the number of parameters it contains. Table~\ref{tab:complexity_comparison_rnnbc} presents a similar comparison of computational complexity as above for modulation classification neural network architectures from the literature. The number of calculations required for a single inference is assumed to be equal to the number of parameters in the model/architecture, and a 64-subcarrier OFDM system is considered. Three models from the current literature are taken into account: VGG, proposed in \cite{68}, ResNet, proposed in \cite{68}, and CNN-AMC, proposed in \cite{69}. The VGG, ResNet, and CNN-AMC models have 257,000, 236,000, and 575,000 parameters, respectively. The modulations considered in VGG and ResNet include OOK, 4ASK, BPSK, QPSK, 8PSK, 16QAM, AM-SSB-SC, AM-DSB-SC, FM, GMSK, and OQPSK, while CNN-AMC considers BPSK, QPSK, 8PSK, 16QAM, 16APSK, 32APSK, and 64QAM. Table~\ref{tab:complexity_comparison_rnnbc} illustrates how combining these models with the RNN-BC approach significantly reduces computational complexity.

\section{CONCLUSION}

The paper presents a novel approach to modulation clas-
sification in OFDM systems, where different modulations
are used in different subcarriers. The proposed approach, which classifies the modulations of N/2 subcarriers using the RNN-BC model, reduces computational complexity by approximately 50 percent compared to running the main classification algorithm (LWNN in this study) N times. This reduction makes the approach more suitable for deployment on edge devices with limited computational power and stringent power constraints.

\bibliographystyle{IEEEtran}
\bibliography{Ref.bib}

\end{document}